\documentclass[twocolumn,showpacs,prb,amsmath,amssymb]{revtex4}
\usepackage{graphicx}
\usepackage{dcolumn}
\usepackage{bm}

\begin{document}
\draft
\title{Heat Conduction and Magnetic Phase Behavior in Electron-Doped Ca$_{1-x}$La$_x$MnO$_3$
($0\leq x\leq 0.2$)}
\author{J.~L.~Cohn$^1$ and J.~J.~Neumeier$^2$}
\affiliation{$^1$ Department of Physics, University of Miami, Coral Gables,
Florida 33124}
\affiliation{$^2$ Department of Physics, Florida Atlantic University, Boca
Raton, Florida}

\begin{abstract}

Measurements of thermal conductivity ($\kappa$) {\it vs}
temperature are reported for a series of
Ca$_{1-x}$La$_x$MnO$_3$($0\leq x\leq 0.2$) specimens. For the
undoped ($x=0$), G-type antiferromagnetic compound a large
enhancement of $\kappa$ below the Ne\'el temperature ($T_N\sim 125
K$) indicates a strong coupling of heat-carrying phonons
to the spin system.  This enhancement
exhibits a nonmonotonic behavior with increasing $x$ and
correlates remarkably well with the small ferromagnetic component of
the magnetization reported previously [Neumeier and Cohn, Phys. Rev. B
{\bf 61} 14319 (2000).]  Magnetoelastic polaron formation appears
to underly the behavior of $\kappa$ and the magnetization
at $x\leq 0.02$.

\end{abstract}
\maketitle

Electronic phase separation has emerged as a paradigm for
describing the ground state of strongly correlated electron
systems.\cite{Dagotto}  It may underly the phenomenon of
colossal magnetoresistance (CMR) in hole-doped
(Mn$^{3+}$-rich) perovskite manganites studied extensively
in recent years. A few studies of
electron-doped (Mn$^{4+}$-rich) manganites
\cite{Earlye-doped} revealed anomalous magnetic properties and drew attention
to these compounds.  Subsequent
investigations\cite{NeumeierCohn,Martin,Savosta,Mahendiran,Santhosh,Respaud,Aliaga}
suggest electron-doped manganites to be novel systems for studies
of phase separation and polaron physics.

\begin{figure}
\vskip .2in
\includegraphics[width=3.5in]{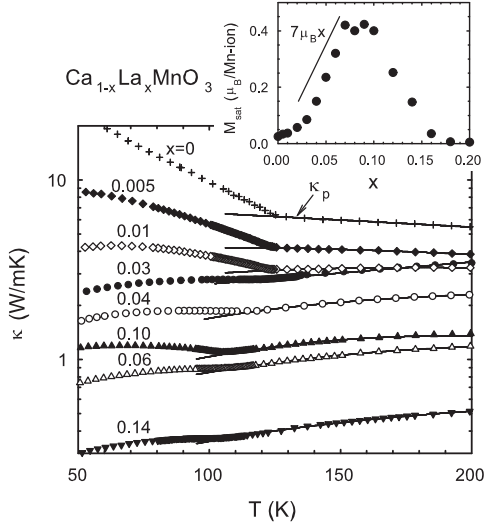}
\vglue -1.3in
\caption{\label{KofT} Thermal conductivity {\it vs} temperature for
Ca$_{1-x}$La$_x$MnO$_3$ polycrystals.  The solid curves are polynomial
fits to data in the paramagnetic phase.  Inset: $T=5$K saturation
magnetization {\it vs} $x$ for Ca$_{1-x}$La$_x$MnO$_3$ polycrystals from
Ref.~\protect\onlinecite{NeumeierCohn}.  The solid line represents
$M_{sat}=7\mu_B$/Mn-ion-electron.}
\end{figure}

Undoped CaMnO$_3$ exhibits G-type antiferromagnetic (AFM) order below
$T_N\approx$125 K.  With $\sim 20\%$ trivalent substitution for
divalent Ca, the system adopts a C-type, orbitally-ordered AFM ground state
with $T_N\sim 150-200$K, depending on the dopant ions.  For Mn$^{3+}$
concentrations between these end points a small ferromagnetic (FM) moment
is observed with $T_C=T_N(G)$ and a nonmonotonic doping behavior; data from
Ref.~\protect\onlinecite{NeumeierCohn} on
Ca$_{1-x}$La$_x$MnO$_3$ ($0\leq x\leq 0.2$) are shown in the inset of
Fig.~\ref{KofT}.

A fundamental issue is whether this FM component reflects a homogeneous
canting of AFM moments for all $x$ in this regime, as originally proposed by
deGennes,\cite{deGennes} or whether some portion of the phase behavior
can be attributed to FM droplets or polarons
as found for the hole-doped compounds.\cite{Hennion}
A very recent neutron scattering study\cite{LingPreprint} of
Ca$_{1-x}$La$_x$MnO$_3$ constrains models for the regime
$0.06\leq x\leq 0.16$: (1) the FM moment is
perpendicular to the G-AFM moment, consistent with homogeneous canting, and
(2) FM clusters with size smaller than $\sim 800${\rm \AA} do not exist.
Similar experiments at low doping $x\leq 0.05$ have not been reported,
but this regime has been investigated theoretically.\cite{Batista,ChenAllen}

The present study of thermal conductivity ($\kappa$) {\it vs}
temperature on Ca$_{1-x}$La$_x$MnO$_3$ ($0\leq x\leq 0.2$) was
motivated by the novel phase behavior of electron-doped
manganites and by prior work\cite{CohnKmang,CohnRev} demonstrating
that the lattice thermal resistivity of manganites is a sensitive
measure of bond disorder arising from distorted MnO$_6$ octahedra.
Electron hopping via double exchange couples the spins to these
octahedral distortions.  For the present system the thermal
conductivity clearly reflects the lattice response to the FM moment
throughout the doping range.  At low doping ($x\leq 0.02$)
the evolution of $\kappa$ and the magnetization suggests a competition
between long-range antiferromagnetism and magnetoelastic polaron formation.

Ca$_{1-x}$La$_x$MnO$_3$ polycrystals were prepared by standard
solid-state reaction; the preparation methods along with
magnetization and resistivity measurements are reported
elsewhere.\cite{NeumeierCohn} Iodometric titration indicated the
oxygen content of all specimens fell within the range 3.00$\pm
0.01$.  The thermal conductivity was measured in a
radiation-shielded vacuum probe using a differential
chromel/constantan thermocouple and steady-state technique.
Typical specimen dimensions were $1\times 1 \times 3$mm$^3$.  Heat
losses via radiation and conduction through leads were
measured in separate experiments and the data corrected
accordingly.  This correction was typically 10-15 \% near room
temperature and $\leq 2$\% for $T\leq 150$ K.  The specimens
have a density of 78$\pm 4\%$ that of fully dense material, with
no systematic dependence on doping; no porosity corrections have
been applied.

Fig.~\ref{KofT} shows $\kappa(T)$ for a series of specimens.  The
data for CaMnO$_3$ ($x=0$) indicate a large enhancement of
$\kappa$ at temperatures below $T_N(G)=130$K (for the remainder of
the paper we write $T_N$ to mean $T_N(G)$). The other compounds
also show an enhancement, but with a diminished magnitude. For all
values of $x$ the electronic contribution to $\kappa$ at $T<150$ K,
as inferred from the electrical
resistivity\cite{NeumeierCohn} and Wiedemann-Franz relation, is
less than 5\% of the measured value. Furthermore, no
substantial changes were observed in resistivities through $T_N$.
Thus the enhancement is associated with either a magnon or phonon
contribution to heat conduction.

\begin{figure}
\vskip -.7in
\includegraphics*[width=3.4in]{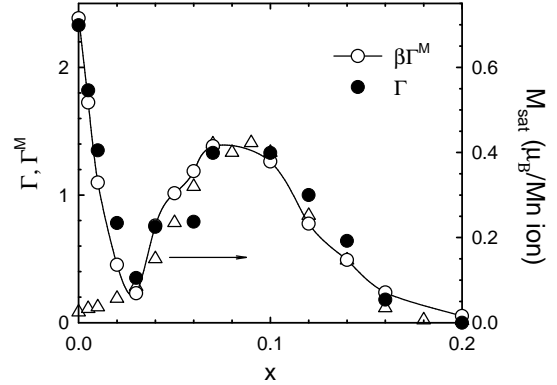}
\vskip -1.4in
\caption{Doping dependence of the dimensionless thermal conductivity
slope change at $T_N$ ($\Gamma$, defined in text), and of the
$T=5$K saturation magnetization from
Ref.~\protect\onlinecite{NeumeierCohn} (open triangles, right
ordinate). Open circles with solid lines are computed from
magnetization data as discussed in the text and Eq. (1).}
\label{Slopes}
\end{figure}

To characterize the enhancement, we define the dimensionless
change in slope, evaluated just below\break
\noindent
$T_N$, as $\Gamma\equiv
-d(\kappa/\kappa_p)/dt|_{t\to 1}$, where $\kappa_p$ is the
$T$-dependent thermal conductivity in the
paramagnetic state ($T>T_N$) and $t=T/T_N$. The behavior of
$\kappa_p$ at $T<T_N$
is taken as the extrapolation of polynomial fits to data at
$T>T_N$ (solid curves in Fig.~\ref{KofT}).  The doping behavior
$\Gamma(x)$ is shown in Fig.~\ref{Slopes}; of particular interest
is the nonmonotonic behavior.  $\Gamma(x)$ appears to be composed
of two contributions: a term strongly decreasing with $x$ and
operative for $x\lesssim 0.02$, and a term proportional to the FM
saturation moment (open triangles and right ordinate,
Fig.~\ref{Slopes}) operative for $x\geq 0.03$.

Increases in $\kappa$ at AF transitions have been observed previously in
MnO\cite{SlackNewman} and LaMnO$_3$\cite{ZhouGoodenough} crystals.
The former material undergoes a substantial crystallographic distortion
below $T_N$ (magnetostriction), and this suggests changes in
the lattice heat conduction as a likely mechanism for its $\kappa$
enhancement.   Lattice anomalies associated with magneto or exchange
striction are quite small
for\cite{RitterNeutron} LaMnO$_3$
and\cite{GranadoRaman,UndopedUltrasound,NeumeierCornelius}
CaMnO$_3$.  Nevertheless, prior work\cite{CohnKmang,CohnRev}
demonstrated that the lattice thermal resistivity of manganites
at $T\leq 300$K is controlled principally by distortions of the MnO$_6$
octahedra through their influence on phonon scattering rates.  The latter
can be substantially more sensitive to internal structural modifications
than are lattice or elastic constants.  Heat conduction by magnons could
contribute to the enhancement, but this seems less likely
given that magnons contribute negligibly to $\kappa$ near
$T_C$ for ferromagnetic compositions.\cite{CohnKmang}  The subsequent
analysis supports a lattice response to magnetic order as the mechanism for
$\Gamma$ in these compounds.

The observation $\Gamma\propto M_{sat}$ for $x\gtrsim 0.03$ is
reminiscent of the behavior found for CMR
compounds.\cite{CohnKmang} The lattice
contribution to $\kappa$ in CMR materials is enhanced below the
zero-field FM transition temperature and in applied magnetic
field at fixed $T$ near $T_C$.
This lattice response correlates with a reduced distortion of the
MnO$_6$ octahedra that accompanies double-exchange mediated charge
delocalization. Underlying the field and $T$-dependent thermal
resistivity ($W=1/\kappa$) of La$_{0.83}$Sr$_{0.17}$MnO$_3$
is a simple magnetization dependence,\cite{CohnKmang}
$W(M)-W(0)\propto -M^2(H,T)$.

We now show that the nonmonotonic behavior of $\Gamma(x)$ for
Ca$_{1-x}$La$_x$MnO$_3$ is remarkably well reproduced by a similar
phenomenological assumption for $T<T_N$: $W(M,T)-W(0,T)\propto M(T)-M_p(T)$.
$M_p(T)$ is the magnetization of the paramagnetic phase.
For zero-field measurements, $M(T)$ is the spontaneous magnetization,
$W(M,T)=\kappa^{-1}$ is the measured
thermal resistivity, and $W(0,T)\equiv \kappa_p^{-1}$ (the hypothetical
thermal resistivity in the absence of magnetic order).
This assumption implies that $\Gamma$ should be proportional to the
normalized change in slope of the magnetization, evaluated just below $T_N$,
\begin{equation}
\label{(1)}
{\Gamma=-\beta {T_N\over M(T_N)}\left( {dM\over dT}
\bigg|_{T\to T_N}-{dM_p\over dT}\bigg|_{T\to T_N}\right)\equiv \beta\Gamma^M}
\end{equation}
\noindent
Using the available $M(T)$ curves\cite{NeumeierCohn} measured at $H=2$kOe,
excellent agreement\cite{noteonEq1} of $\Gamma(x)$ with Eq.~(1)
was found for all $x$, with proportionality constant
$\beta=2.52\times 10^{-4}$ (open circles and
solid curve, Fig.~\ref{Slopes}).

This result is most easily interpreted in the regime $x\gtrsim
0.03$ where $\Gamma\propto M_{sat}$.  Though no
insulator-metal transition takes place at $T_N$ akin to
that at $T_C$ in CMR compounds, there is clear evidence in
electrical resistivity measurements near $x=0.10$ that some
electron delocal-\break
\begin{figure}
\includegraphics*[width=3.25in]{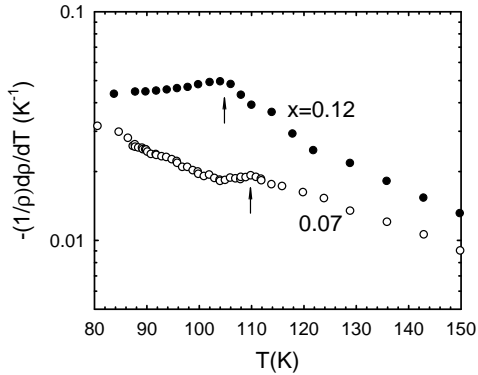}
\vglue -1.7in
\caption{Temperature derivative of electrical resistivity {\it vs}
temperature for Ca$_{1-x}$La$_x$MnO$_3$ specimens with $x$ near 0.10.  Features
marked by arrows indicate electronic delocalization below $T_N\approx 110$K.}
\label{deloc}
\end{figure}
\noindent
ization takes place below $T_N$; the slopes,
$-(1/\rho)d\rho/dT$, exhibit an abrupt change at the transition
(Fig.~\ref{deloc}).  In analogy with the case of
CMR materials, the correlation between $\Gamma$ and magnetization
in this regime is plausibly attributed to enhanced electron hopping
mediated by double-exchange between aligned (FM droplet scenario)
or partially aligned (canting scenario) spins.  Enhanced electron transfer
reduces the average distortion of the MnO$_6$ octahedra
and associated phonon scattering within the FM regions of the specimen.
That $\Gamma$ follows both $M_{sat}$ and $\Gamma^M$ is consistent
with a conventional magnetization of the form $M=M_{sat}f(T)$
where $f(T)$ reflects the order parameter of the FM phase.

\begin{figure}
\includegraphics*[width = 3.25in]{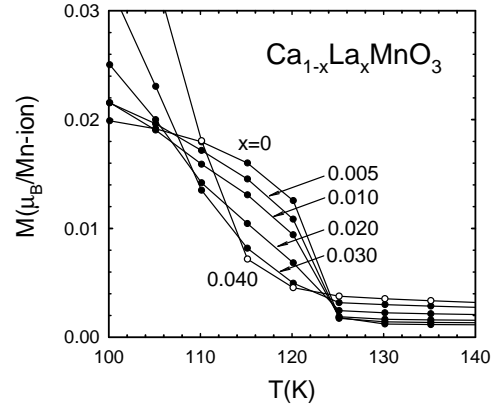}
\vglue -1.7in
\caption{Magnetization at $H=2000$Oe for lightly-doped specimens.}
\label{LowxMofT}
\end{figure}

The regime $x\leq 0.02$ is more complicated since
$M_{sat}$ has a very different $x$ dependence from
that of $\Gamma^M$ (and $\Gamma$).  These different doping
behaviors entail a crossing of the $M(T)$
curves for different $x$ at $T<T_N$ (Fig.~\ref{LowxMofT}).  The
data suggest that two independent
components contribute to the magnetization in this regime, one
with $T_C=125$K, characterizing the undoped specimen, and the other
with $T_C\lesssim 115$K, characterizing $x\geq 0.04$.  A smooth
evolution between the two is reflected in the $M(T)$ data for the
intervening compositions.  This coexistence is most evident in the curve for
$x=0.02$ as an inflection near $T=110$ K.

$\Gamma$ follows the diminution of the higher-$T$
transition and presumably has its origin in the same coupling
between spins and octahedral distortions underlying the
response at $x\geq 0.03$.  Supporting this hypothesis
is a recent study of Raman scattering in similar compounds.\cite{GranadoRaman}
Sharply enhanced Raman intensities
at $T<T_N$ for low-frequency, rotational and bending modes of the
oxygen octahedra were observed for CaMnO$_3$.  With increasing
La doping this enhancement below $T_N$ was diminished, and
was absent for $x$=0.03, very similar to the trend observed here for $\Gamma$.
It seems likely that the two phenomena are related.

Another important experimental result relevant\break
\noindent
to a description
of the data at low $x$ comes from recent magnetic neutron
scattering studies of an $x=0.02$ compound in magnetic
field.\cite{GranadoPrivate} The field dependence of the
AF scattering intensity was inconsistent with a FM component
arising from either uniform spin canting
or ferrimagnetism; the FM component is {\it decoupled} from the
AF background in applied field.

These observations implicate magnetoelastic polarons in the FM and
lattice response at low $x$.  In this regard the behavior of the
slope $dM_{sat}/dx$ is of interest (inset, Fig.~\ref{KofT}).  It
increases from $\sim 1\mu_B$/Mn-ion-electron for $x\leq 0.02$ to
$\sim 7\mu_B$/Mn-ion-electron for $0.03\leq x\leq 0.08$.
The latter is the value expected if each La dopant
adds a symmetric 7-site FM polaron,\cite{NeumeierCohn} determined in
calculations\cite{ChenAllen} to be the stable ground state for this system.
The value $x\simeq 0.03$ appears to mark a crossover between
regimes.  The mean spacing between dopant ions is estimated as,
$r_{La}=(3V/16\pi x)^{1/3}$ ($V=207{\rm\AA}^3$ is the unit
cell volume,\cite{LingPreprint} containing 4 f.u.'s).  For $x=0.03$,
$r_{La}\simeq 7.4{\rm\AA}$, equal to the 3rd-nearest neighbor Mn
distance.  This would be the distance between the centers of
7-site polarons sharing a single Mn site (a polaron ``cluster''),
but other polaron configurations are close in ground-state
energy and might be favored if interactions
not yet considered in calculations\cite{ChenAllen} (e.g. defects,
next-nearest neighbor exchange) play a role.  For example, larger
polarons with canted-spin arrangements are interesting candidates because
the large-scale clusters anticipated at higher doping
could evolve smoothly into the long-range, spin-canted state proposed
on the basis of neutron scattering\cite{LingPreprint} for $x\geq 0.06$.

A model that distinguishes between isolated and clustered polarons
describes the qualitative features of the experimental results.
We propose that the $T_C=125$ K transition in Fig.~\ref{LowxMofT} is
associated with isolated polarons (i.e. those without adjacent polarons)
and the lower-$T_C$ transition to clusters of two or more polarons.
This is plausible since the balance between kinetic, lattice and spin
energies, possibly different for a cluster, could yield a polaronic state
less robust against thermal fluctuations.
Presumably a small (principally isolated) polaron density at $x=0$ is
associated with native defects (e.g. oxygen or Mn vacancies).
With increasing La doping, polaron clusters are produced at the expense
of isolated polarons.  For a random distribution, the probability for two or
more polarons to share a Mn site grows rapidly for 7-site or larger polarons
given the large number of Mn sites over which electrons may be distributed
to define a cluster.  This provides a natural explanation for the rapid
decrease of $\Gamma$ and $\Gamma^M$ with $x$ at $x\leq 0.02$, and for their
minima at $x\approx 0.03$ where the FM transition is maximally
nonuniform.

What is the origin of the small value for $dM_{sat}/dx$
at $x\leq 0.02$ and its increase near $x\simeq 0.03$?  One
possibility is that these features reflect differing magnetic
moments for isolated and clustered polarons (e.g. different canting
angles for spin-canted polarons).  Alternatively,
these features might reflect a change in the number of polarons induced per
dopant, e.g. if perturbations associated with defects (vacancies, La ions)
suppress polaron formation at $x\leq 0.02$.  Experimental
investigations that better define the polaron
characteristics are required to refine these ideas.

In summary, systematic changes in the behavior of the thermal conductivity
at the AF transition in La-doped CaMnO$_3$ and their
correlation with magnetization measurements indicate that the lattice
thermal resistivity is a sensitive probe of FM interactions through the
coupling of spins to local distortions of the MnO$_6$ octahedra.  This
extends similar conclusions obtained previously for hole-doped, CMR
compounds\cite{CohnKmang,CohnRev} to the present system where the ground
state appears to consist of magneto-elastic polarons ($x\leq 0.02$) and
a spin-canted phase ($x\geq 0.06$).  The crossover between
these two regimes, clearly manifested in both the thermal conductivity
and magnetization data, appears to reflect novel polaron physics
and is of particular interest for further study.

The authors acknowledge experimental assistance from Dr. B. Zawilski and
Dr. R. Maier.  The work at the University of Miami was supported, in part, by NSF
Grant No.'s DMR-9631236 and DMR-0072276, and at Florida Atlantic University by
DMR-9982834.

\end{document}